\documentclass[11pt]{JHEP3} 

\usepackage{epsfig,multicol,bbm}

\newcommand\fverb{\setbox\fverbbox=\hbox\bgroup\verb}
\newcommand\fverbdo{\egroup\medskip\noindent%
            \fbox{\unhbox\fverbbox}\ }
\newcommand\fverbit{\egroup\item[\fbox{\unhbox\fverbbox}]}
\newbox\fverbbox

\usepackage{amsmath}
\usepackage{amssymb}
\usepackage{amsfonts}
\usepackage{latexsym}
\usepackage{graphicx}
\usepackage{dcolumn}
\usepackage{bm}
\usepackage{empheq}

\usepackage{times}  
\newcommand{\eq}[1]{\begin{equation}#1\end{equation}}

\newcommand{\ea}[1]{\begin{equation}\begin{aligned}#1\end{aligned}\end{equation}}
\newcommand{\itm}[1]{\begin{itemize}#1\end{itemize}}
\newcommand{\od}[2]{\frac{\textrm{d} #1}{\textrm{d} #2}}  
\newcommand{\lrp}[1]{\left( #1 \right)}  
\newcommand{\lrsb}[1]{\left[ #1 \right]}  
\newcommand{\lrcb}[1]{\left\{ #1 \right\}}  
\newcommand{\lrab}[1]{\left\langle #1 \right\rangle}  
\newcommand{\lrmb}[1]{\left| #1 \right| } 
\newcommand{\soev}[3]{\lrab{ #1 \lrmb{ #2}#3 }} 
\def\la{\langle}
\def\ra{\rangle}

\def\rd{\partial}
\def\vk{\bm{k}}
\def\vp{\bm{p}}

\def\dep{\delta\phi}
\def\dop{\dot{\phi}}

\def\qsg{Q_{\sigma}}
\def\qs{Q_{s}}

\def\I{\mathrm{I}}

\title{Loop Corrections to Cosmological Perturbations in Multi-field Inflationary Models}

\author{Xian Gao\\
    Key Laboratory of Frontiers in Theoretical Physics,\\
     Institute of Theoretical Physics, Chinese Academy of
    Sciences, Beijing 100190, P.R.China\\
    E-mail: \email{gaoxian@itp.ac.cn}
    }

\author{Fanrong Xu\\
    Key Laboratory of Frontiers in Theoretical Physics,\\
     Institute of Theoretical Physics, Chinese Academy of
    Sciences, Beijing 100190, P.R.China,\\ and \\
      Maryland Center for Fundamental Physics and  Department of Physics , \\
  University of Maryland, College Park, MD 20742, USA\\
    E-mail: \email{frxu@itp.ac.cn}
     }

\preprint{  CAS-KITPC/ITP-109}  

\abstract{
    We investigate one-loop quantum corrections to the power spectrum of
    adiabatic perturbation from entropy modes/adiabatic mode
    cross-interactions in multiple DBI inflationary models. We find that due to the
non-canonical kinetic term in DBI models, the loop corrections are
enhanced by slow-varying parameter $\epsilon$ and small sound speed
$c_s$. Thus, in general the loop-corrections in multi-DBI models can
be large. Moreover, we find that the loop-corrections from
adiabatic/entropy cross-interaction vertices are IR finite. }


\begin{document}


\section{Introduction}

Current observational data support standard $\Lambda$CDM model
greatly \cite{Komatsu:2008hk}, in which primordial perturbations
which are assumed to be responsible for Cosmic Microwave Background
anisotropies and large-scale structure formation are generated from
quantum fluctuations and stretched to superhorizon scales during
inflation (see e.g. \cite{Lyth:1998xn} for a review). Standard
single-field slow-roll inflationary models predict almost
scale-invariant, Gaussian and adiabatic primordial fluctuation.

Actually, inflation itself is not a single model, but rather a
theoretical framework. From the point of view of power spectrum,
many inflation models are ``degenerate". Power spectrum, or strictly
speaking, tree-level two-point correlation functions of cosmological
perturbation do not give us an unique theory of inflation. In fact,
phenomena beyond linear-order have been extensively investigated
over the past several years.

The most significant progress beyond power spectrum is the
investigation of statistical non-gaussianities of CMB anisotropies
and primordial fluctuations (see e.g. \cite{Bartolo:2004if} for a
nice review). From the field theoretical point of view,
non-gaussianity describes the \emph{interactions} of perturbations,
which will cause non-vanishing higher-order correlation functions.
Such interactions are mandatory in any realistic inflationary
models, which come from both the non-linear nature of gravity and
the self-interactions of inflation field(s). In standard slow-roll
inflation scenario, however, the non-gaussianities have been proved
too small to be detected \cite{Maldacena:2002vr}, even with PLANCK.
Thus, any detection of non-gaussianties would not only rule out the
simplest inflation models but also give us valuable insight of
fundamental physics of inflation. Various models have been
investigated to generate large non-gaussianties by introducing more
complicated kinetic terms
\cite{Seery:2005wm,Chen:2006nt,Huang:2006eh,Arroja:2008ga} or more
fields
\cite{Seery:2005gb,Langlois:2008qf,Langlois:2008wt,Easson:2007dh,Huang:2007hh,Arroja:2008yy,Cai:2008if,Cai:2009hw,Huang:2009vk,Huang:2009xa,Gao:2008dt,Gao:2009gd}.

Interactions not only cause non-vanishing higher-order correlation
functions which are responsible for non-gaussianities of primordial
perturbations, they also cause \emph{quantum} loop-corrections. It
is possible that such loop corrections may be large in their own
right. Especially, we could expect large loop corrections in models
which have also large non-gaussianties, since both of them describe
the interactions among perturbations. Moreover, by collecting
signatures of both quantum loop corrections and non-gaussianties, we
will obtain a more sensitive test of the physics of inflation.

Early estimations of loop effects in cosmological perturbations were
pioneered by Mukhanov, Abramo and Brandenberger
\cite{Mukhanov:1996ak,Abramo:1997hu}, Abramo and Woodard
\cite{Abramo:1998hi}, and Unruh \cite{Unruh:1998ic}. Subsequently,
loop effects have been investigated by many authors
\cite{Onemli:2002hr,Prokopec:2002uw,Prokopec:2003bx,Onemli:2004mb,Brunier:2004sb,Kahya:2005kj,Kahya:2006ui,Kahya:2009sz,Brandenberger:2002sk,Geshnizjani:2003cn,Brandenberger:2004ki,Brandenberger:2004kx,Brandenberger:2004ix,Martineau:2005aa,Gerstein:1971fm,Boyanovsky:2005sh,Boyanovsky:2005px,Chaicherdsakul:2006ui,Losic:2005vg,Kahya:2006hc,Bilandzic:2007nb,Kim:2007sb}.
Recently, stimulated by works of S. Weinberg
\cite{Weinberg:2005vy,Weinberg:2006ac}, loop corrections to
primordial fluctuations have been re-investigated by several authors
\cite{Sloth:2006az,Sloth:2006nu,Seery:2007we,Seery:2007wf,Dimastrogiovanni:2008af,Adshead:2008gk,Adshead:2009cb,Urakawa:2008rb}.
It has been found that cosmological loop corrections seem to have
some infrared divergences, which scale as $\ln (k\ell)$, where
$\ell^{-1}$ is some IR comoving momentum cut-off
\cite{Sloth:2006az,Sloth:2006nu,Seery:2007we,Seery:2007wf}. Other
discussions on cosmological loop corrections and the infrared
divergences can be found in
\cite{Lyth:2007jh,Bartolo:2007ti,Enqvist:2008kt,Riotto:2008mv,Urakawa:2009my,Urakawa:2009gb,Janssen:2008px}.
The presence of IR divergences implies that loop-contributions may
potentially give large corrections to tree-level results, although
it was argued in \cite{Lyth:2007jh,Bartolo:2007ti,Enqvist:2008kt}
that these IR divergent corrections do not contribute in quantities
that are directly observable.

Actually, now there are two types of so-called ``loop corrections"
in the literatures. One type is the quantum loops (q-loop) which
originate from the interactions among fields and are evaluated using
standard quantum field theory, while the other type is the classical
loops (c-loop) which arise mathematically from some ``loop-type
integrals" due to the non-linear map from $\dep$ to $\zeta$
\cite{Cogollo:2008bi,Rodriguez:2008hy,Byrnes:2008zy}. In this work,
we focus on the quantum loops. However, it is well-known that the
observational variable is the curvature perturbation $\zeta$ and
there has non-linear relation between $\zeta$ and $\delta$ during
superhorizon evolution (e.g. in terms of the ``$\delta N$-formalism"
\cite{Starobinsky:1986fxa,Sasaki:1995aw,Lyth:2004gb}). Thus, these
``loop" effects also have to be treated seriously ,and especially
they can be significant, if the higher-order derivatives of
e-folding number N with respect to the inflaton fields are extremely
large. While in \cite{Byrnes:2008zy} it was shown that the loop
corrections for $\zeta$ are always suppressed during inflation, if
one assume that the fields follow a classical background trajectory.

 Up to now, one-loop
corrections to the power spectrum of the cosmological perturbations
have been investigated in single scalar-field loops
\cite{Sloth:2006az,Sloth:2006nu,Seery:2007we,Seery:2007wf,Weinberg:2005vy,Weinberg:2006ac},
 concerning effects from many light scalar fields \cite{Adshead:2008gk},
and from graviton loops \cite{Dimastrogiovanni:2008af}. Actually, it
has been recognized that it is hard to generate large ``local-type"
non-gaussianities in single-field models, unless we abandon the
slow-roll-type conditions. Thus, the investigation of multiple field
models has particular importance. In multiple field inflationary
models, perturbations can be decomposed instantaneously into one
adiabatic mode and several entropy modes \cite{Gordon:2000hv} (see
\cite{Bassett:2005xm} for a review). There are interactions between
adiabatic mode and entropy modes.

In this work, we investigate the one-loop corrections to adiabatic
power spectrum due to the interactions between entropy mode and
adiabatic mode, at the horizon exiting. We find that due to the
non-canonical kinetic term in DBI models, the loop corrections are
enhanced by slow-varying parameter $\epsilon$ and small sound speed
$c_s$, as the enhancement of non-gaussianties in such models. Thus,
in general, the loop-corrections in multi-DBI models can be large.
Moreover, we find that the loop-corrections from adiabatic/entropy
cross-interaction vertices are IR finite. This is because in the
limit of speed sound $c_s$, the leading-order three-point
interactions are totally ``derivatively interacted". While the
derivatives (both with respect to time and space) give momentum
factors which cancel the momentum factors in the denominator from
the tree-level correlation functions to make the loop-integrals IR
safe.

Quantum cosmological correlation functions are evaluated by using
``in-in formalism" (also dubbed as ``Schwinger-Keldysh formalism" )
\cite{Schwinger:1960qe,Calzetta:1986ey,Jordan:1986ug} (also see e.g.
\cite{Weinberg:2005vy} for a nice review), which we also give a
brief review in Appendix A. In \cite{Adshead:2009cb} the application
of ``in-in" formalism on cosmological perturbation, especially on
loop corrections, has been investigated in details.

This paper is organized as follows. In section 2, firstly we briefly
review the linear perturbation of multiple-DBI models, then we
describe the third-order perturbation action, especially the
three-point adiabatic/entropy modes cross-interaction vertices. In
section 3, we calculate the one-loop corrections to the adiabatic
power spectrum from the adiabatic/entropy modes cross-interaction
vertices, by using the ``in-in" formalism. Finally, we make a
conclusion and discuss the related issues.

In this paper, we set $\hbar = c = 8\pi G=1$.

\section{Basic Setup}

Perturbation theory of general multiple field modes with
non-canonical kinetic terms up to third-order action has been
calculated by several authors
\cite{Langlois:2008qf,Langlois:2008wt,Easson:2007dh,Huang:2007hh,Arroja:2008yy,Cai:2008if,Cai:2009hw,Ji:2009yw,Gao:2008dt}.
In this work, we focus on multi-field DBI models
\cite{Langlois:2008qf,Langlois:2008wt,Easson:2007dh,Huang:2007hh,Arroja:2008yy,Cai:2008if,Cai:2009hw}.
In this section, we briefly review previous known results and setup
our conventions.

\subsection{The model and the background}

In this work, we focus on two-field DBI inflationary models, with
action of the form
    \eq{
        S = \int d^4x\, \lrp{ \frac{1}{2}R + \mathcal{L}^{\phi}}\,,
    }
with{\footnote{In general, the full DBI-type action takes the form
$\mathcal{L} = -\frac{1}{f} \lrp{ \sqrt{D} - 1 } -V$, with
    \[
        D = \det\lrp{ \delta_{IJ} - 2fX_{IJ} } \equiv 1 -2f X + 4f^2
        X^{[I}_I X^{J]}_J - 8f^3 X^{[I}_{I} X^{J}_{J} X^{K]}_{K} +
        16 f^4 X^{[I}_{I} X^{J}_{J} X^{K}_{K} X^{L]}_{L} \,.
    \]
However, it is shown that higher-order terms proportional to $f^3$
and $f^4$ do not contribute to the calculation of power spectra and
leading order bispectra. Thus, in this work we neglect these terms
for simplicity. One can refer
\cite{Langlois:2008qf,Langlois:2008wt,Easson:2007dh,Huang:2007hh,Arroja:2008yy,Cai:2008if,Cai:2009hw}
for detailed investigations of multi-DBI inflation models.}}
    \eq{
        \mathcal{L}^{\phi}  \equiv
        -\frac{1}{\sqrt{f(\phi^I)}} \lrp{ \sqrt{ 1- 2fX + 2f^2 \lrp{ X^2 - X_{IJ}X^{IJ} } } } - V(\phi^I)
        \,,
    }
where
    \eq{
        X^{IJ} \equiv -\frac{1}{2} \rd_{\mu} \phi^I \rd^{\mu} \phi^J
        \,,
    }
and $X\equiv G_{IJ}X^{IJ}$.

The slow-rolling parameter is defined as usual
    \eq{
        \epsilon = -\frac{\dot{H}}{H^2}\,.
        }
The background ``inflaton velocity" $\dot{\sigma}$ can be related
with $\epsilon$ as
    \eq{{\label{bg_inf_vel}}
        \dot{\sigma} \equiv \sqrt{2X} = H \sqrt{2\epsilon c_s} \,,
    }
where $c_s$ plays the role of the propagating speed of the
perturbation. In DBI inflation model,
    \eq{
        c_s^2 \equiv 1- 2f X \,,
    }
where $X$ is evaluated with the background quantities.

\subsection{Brief review of linear perturbation}

In multi-field inflation models, perturbations can be
instantaneously decomposed into one adiabatic mode and several
entropy modes \cite{Gordon:2000hv}. Restricting to two-field case,
we introduce the basis $\lrcb{e_{\sigma}, e_s}$ with $e_{\sigma}^I
\equiv \dop^I/\dot{\sigma}$, where $\dot{\sigma}$ is the background
inflaton velocity defined in (\ref{bg_inf_vel}), and $e_s^I$ is
orthogonal to $e^I_{\sigma}$. Then we introduce the decomposition
    \eq{
        Q^I \equiv \delta\phi^I = \qsg e^I_{\sigma} + \qs e^I_s \,,\qquad\qquad I=1,2 \,.
    }
$Q_{\sigma}$ and $Q_s$ are adiabatic mode and entropy mode
respectively.

After the adiabatic/entropy modes decomposition, the second-order
perturbation action for two-field DBI inflation model up to
leading-order was derived in
\cite{Langlois:2008qf,Langlois:2008wt,Arroja:2008yy} in
spatially-flat gauge:
    \eq{\label{2rd_aciton}
        S_{(2)} = \int~d\eta d^3x~\frac{a^2}{2 c^3_s} \left\{ \left[(\qsg')^2-c^2_s(\rd_i \qsg)^2\right]
        +c_s^2 \left[(Q'_s)^2-c^2_s(\rd_i \qs)^2\right]\right\}\,,
        }
where $H\equiv \dot{a}/a$ is the Hubble parameter, $\eta$ is the
comoving time, and a prime denotes the derivative with respect to
$\eta$. In deriving (\ref{2rd_aciton}), we assume that $c_s$ is
approximately constant and neglect coupling between adiabatic and
entropy modes which may give significant effect in some cases, and
keep only the leading order terms. Note that in general multi-field
models, adiabatic mode and entropy modes can propagate with
different (and arbitrary) speeds of sound, which was first pointed
out in \cite{Easson:2007dh,Huang:2007hh} (see also
\cite{Arroja:2008yy,Cai:2008if,Cai:2009hw,Ji:2009yw,Gao:2008dt}).
Multi-field DBI model is a special case, where adiabatic mode and
entropy modes propagate with the same speed of sound $c_s$.

In canonical quantization procedure, we write \ea{
        \qsg(\eta, \vk) &= u_k(\eta) a_{\vk} + u^{\ast}_k(\eta)
        a^{\dag}_{-\vk} \,,\\
        \qs(\eta, \vk) &= v_k(\eta) a_{\vk} + v^{\ast}_k(\eta)
        a^{\dag}_{-\vk} \,.
    }
In de Sitter approximation $a(\eta)=-1/H\eta$, the mode solutions
can be easily solved \ea{{\label{mode_function}}
    u_k (\eta) &= \frac{ H}{\sqrt{2k^3}} \lrp{1+i c_sk \eta } e^{-i c_sk\eta}  \,,\\
    v_k (\eta) &= \frac{ H}{c_s\sqrt{2k^3}} \lrp{1+i c_sk \eta } e^{-i
    c_sk\eta} \,.
} Note that this choice of mode functions corresponds to the
Bunch-Davis vacuum.
 Now it is straightforward to write down the tree-level two-point
correlation functions (see Appendix A for more details),
    \ea{\label{2pf_tree}
        \soev{0}{ Q_{\sigma}(\eta,\vk_1) Q_{\sigma}(\eta',\vk_2) }{0}
         &= (2\pi)^3\delta^3(\vk_1 + \vk_2)\, G_{k_1}(\eta,\eta') \;,\\
        \soev{0}{ Q_{s}(\eta,\vk_1)Q_{s}(\eta',\vk_2) }{0}
        &= (2\pi)^3\delta^3(\vk_1+ \vk_2)\,  F_{k_1}(\eta,\eta') \;,\\
        }
where{\footnote{In the literatures on ``in-in formalism", it  used
to define ``two" types of Green's functions: $G^>_k(\eta,\eta')
\equiv u_k(\eta_1) u^{\ast}(\eta_2)$ and $G^<_k(\eta,\eta') \equiv
u^{\ast}_k(\eta_1) u(\eta_2)$, keeping $\eta >\eta'$. It is easy to
see that, mathematically, $G^<_k(\eta,\eta') \equiv
\lrp{G^>_k(\eta,\eta')}^{\ast} \equiv G^>_k(\eta',\eta)$. Thus, if
we relax the artificial restriction $\eta >\eta'$, there is no need
to keep track of this odd notation. In this paper, we use
$G_k(\eta,\eta')$ for simplicity.}}
    \eq{ \label{greenfunction}
        G_k(\eta,\eta') \equiv u_k(\eta)u^{\ast}_k(\eta')\,,\qquad\qquad F_k(\eta,\eta')
        \equiv
        v_k(\eta)v^{\ast}_k(\eta')\,,
        }
are sometimes called Wightman functions.

Then the power spectra for $Q_{\sigma}$ and $Q_{s}$ are
    \ea{\label{power sigma s}
        P_{\sigma} (k) &= |Q_{\sigma^{\ast}}|^2=\frac{H_{\ast}^2}{2k^3}\;,\\
        P_{s} (k) &= |Q_{s^{\ast}}|^2=\frac{H_{\ast}^2}{2k^3c^2_s}\;,
        }
respectively, where the subscript $\ast$ indicates that the
corresponding quantity is evaluated at the sound horizon crossing
$kc_s=aH$.


\subsection{Adiabatic-entropy modes three-point cross-interactions}

The third-order perturbation action of two-field DBI models was
derived in \cite{Langlois:2008qf,Langlois:2008wt} (see also
\cite{Easson:2007dh,Huang:2007hh,Arroja:2008yy}). In the
leading-order of slow-varying parameters and in the limit of small
$c_s$, we have
    \ea{ \label{3rd_action}
        S_3
        &= \int~d\eta d^3x\, \frac{a}{2 c^3_s \dot{\sigma} } \left\{ c^{-2}_s
        \left[(Q'_{\sigma})^3 - c^2_s Q'_{\sigma}(\nabla Q_{\sigma})^2\right]\right. \\
        &\qquad \left. + \left[ Q'_{\sigma}(Q'_s)^2
        + c^2_s Q'_{\sigma}(\nabla Q_s)^2 - 2c^2_s Q'_s \nabla Q_s \nabla
        Q_{\sigma} \right]\right\} \,.
        }
The first line in (\ref{2rd_aciton}) is $\qsg$ three-point
self-interaction terms, while the second line is $\qsg$-$\qs$
cross-interaction terms. Note that there is no $\qs$
self-interaction term.

In this work, we focus on the cross-interaction terms involving
entropy mode and adiabatic mode. The interacting Hamiltonian at
third-order is given by $H_{\textrm{i}} = -L_{(3)}$, where $L_{(3)}$
is the Lagrangian read from (\ref{3rd_action}). There are three
cross-interaction vertices, \eq{{\label{H_abc}}
    H_{\textrm{i}} = H_{\textrm{i}a} + H_{\textrm{i}b} +
    H_{\textrm{i}c} \,,
} where subscript ``i" denotes interacting part of the Hamiltonian,
and \ea{{\label{3p_vertex}}
    H_{\textrm{i}a}(\eta) &= \frac{g}{ \eta }
    \int \prod^3_{i=1}  \frac{d^3k_i}{(2\pi)^3}\,
    (2\pi)^3\delta^3( \vk_{123} )\, Q'_{\sigma}(\eta,{\vk}_1)\, Q'_{s}(\eta,\vk_2 )\, Q'_{s}(\eta,\vk_3) \,,\\
    H_{\textrm{i}b}(\eta) &= - \frac{g\,c_s^2}{ \eta }  \int
    \prod^3_{i=1}\frac{d^3k_i}{(2\pi)^3}\,
    (2\pi)^3\delta^3(\vk_{123} )\, (\vk_2 \cdot \vk_3)\, Q'_{\sigma}(\eta,\vk_1)\, Q_{s}(\eta,\vk_2)\, Q_{s}(\eta,\vk_3) \,,\\
    H_{\textrm{i}c}(\eta) &= \frac{2g\,c_s^2}{ \eta } \int \prod^3_{i=1}
    \frac{d^3k_i}{(2\pi)^3} \,
    (2\pi)^3\delta^3(\vk_{123} )\, (\vk_1 \cdot \vk_2)\,
    \qsg(\eta,\vk_1)\,  Q_{s}(\eta,\vk_2)\, Q'_{s}(\eta,\vk_3)\,,
    }
where we have changed into momentum space and
\eq{{\label{eff_coupling}}
        g \equiv \frac{1}{ 2H^2 c^3_s \sqrt{2\epsilon c_s} } \,,
    }
which plays the role of an ``effective coupling constant" and we
assume to be approximately constant during inflation. In deriving
(\ref{3p_vertex}) we have used de Sitter approximation $a=-1/H\eta$,
and $\vk_{123} \equiv \vk_1 +\vk_2 +\vk_3$ for simplicity.

\begin{figure}[h]
    \centering
    \begin{minipage}{0.8\textwidth}
    \centering
    \begin{minipage}{0.3\textwidth}
    \centering
    \includegraphics[width=2.5cm]{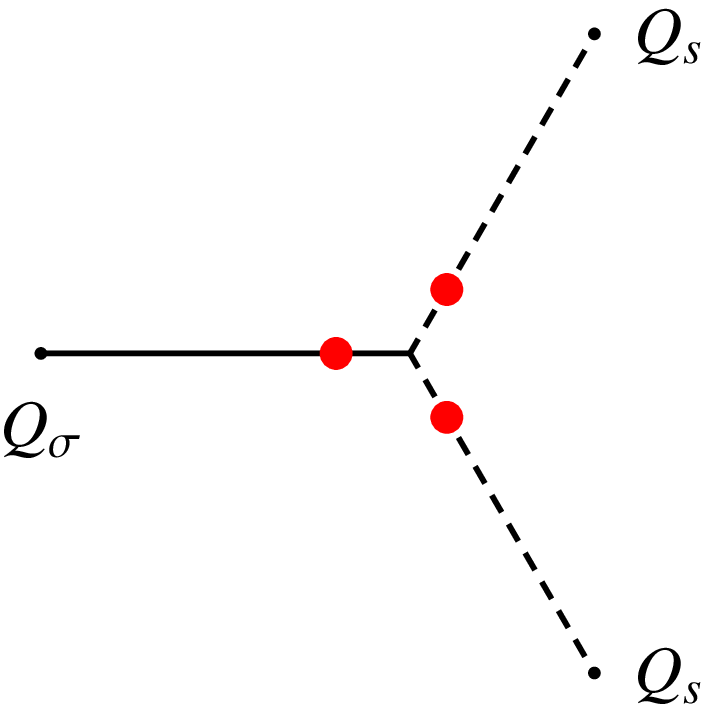}
    \end{minipage}
    \begin{minipage}{0.3\textwidth}
    \centering
    \includegraphics[width=2.5cm]{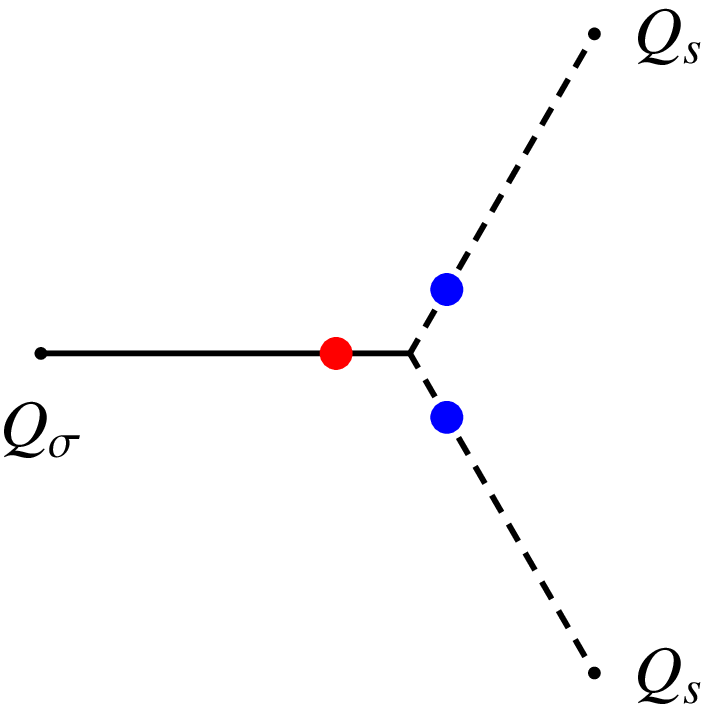}
    \end{minipage}
    \begin{minipage}{0.3\textwidth}
    \centering
    \includegraphics[width=2.5cm]{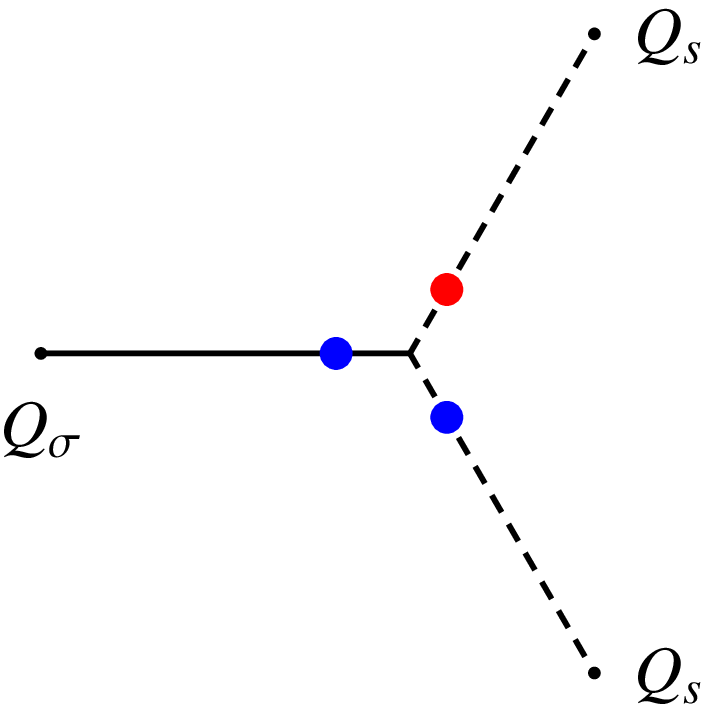}
    \end{minipage}
    \caption{Diagrammatic representations of three types of adiabatic-entropy three-point cross-interaction vertices. A red dot denotes temporal  derivative while a blue dot denotes spatial derivative, or in fourier space, a momentum factor.}
    \label{fig_vertices}
    \end{minipage}
    \end{figure}

From (\ref{3p_vertex}) it can be seen that the three types of
three-point interaction vertices are all ``derivative interactions",
as shown in Fig.\ref{fig_vertices}. This is different from ordinary
field theory where fields are locally coupled. Moreover, as we will
see, the derivatives bring momentum factors, which will cancel the
momentum factors in the denominator come from the two-point
correlation functions, and increase the finiteness in IR.
Especially, in this work we will show that loop corrections arising
from the above three-point vertices are all IR finite.

There is another issue should be emphasized. In standard slow-roll
inflation models, the ``effective couplings" of the interaction
vertices are greatly suppressed by slow-roll parameters (this is
also why primordial non-gaussianies are much small in these models)
and thus contributions to the correlation functions from
multi-vertices diagrams are suppressed and can be neglected at least
in the leading-order. However, this is no longer the case for $c_s
\ll 1$ models, where the strength of the interactions are much
larger. This can be seen directly from (\ref{eff_coupling}), the
effective coupling is enhanced by small $c_s$. For example, in
\cite{Gao:2009gd}, trispectrum from exchanging scalar modes were
investigated, it was shown that in $c_s \ll 1$ models, in general
the contribution from two-vertices diagrams cannot be neglected in
comparison with one-vertices diagrams. The essential idea is that,
contributions from diagrams with more numbers of vertices are of the
same importance in comparison with contributions from diagrams with
less numbers of vertices, even in the leading-order. This is also
one of the motivations of this work.

\section{One-loop Entropy Mode Corrections}

Correlation functions in cosmological context are calculated by
using ``in-in" formalism
\cite{Schwinger:1960qe,Calzetta:1986ey,Jordan:1986ug} (see Appendix
A for a brief review). In this work, we focus on the one-loop
contributions arising from the three-point vertices, thus we need
\eq{{\label{master}}
            \begin{aligned}
                \lrab{ \qsg(\eta_{\ast}, \vk_1) \qsg(\eta_{\ast}, \vk_2 ) }_{1-\textrm{loop}} &= - 2\, \textrm{Re} \lrsb{ \int^{\eta_{\ast} }_{{ -\infty^+} } d\eta_1\, \int^{\eta_1}_{{ -\infty^+}
        } d\eta_2\, \soev{0}{ \qsg(\eta_{\ast}, \vk_1) \qsg(\eta_{\ast}, \vk_2 ) \,  H_{\textrm{i}}(\eta_1)\, H_{\textrm{i}}(\eta_2) }{0} } \\
        &\qquad\qquad + \int^{\eta_{\ast} }_{{ -\infty^-} } d\eta_1 \int^{\eta_{\ast}}_{{ -\infty^+} }
        d\eta_2\, \soev{0}{ H_{\textrm{i}}(\eta_1) \, \qsg(\eta_{\ast}, \vk_1) \qsg(\eta_{\ast}, \vk_2 ) \,
         H_{\textrm{i}}(\eta_2) }{0} \,,
         \end{aligned}
            }
where we have written explicitly the lower bound for
time-integrals{\footnote{One can refer a recent work
\cite{Adshead:2009cb}, for a detailed discussion on the application
of ``in-in" formalism on cosmological correlation functions, and the
issue of the lower bound of time integrals.}}: $-\infty^\pm \equiv
-\infty (1\mp i\epsilon)$ (see Appendix A for details). Note that
all quantities on the right-hand side in (\ref{master}) are
``interaction-picture" quantities. In this paper we choose
$H_{\textrm{i}}$ involving three-point cross-interactions of $\qsg$
and $\qs$ which is given by (\ref{H_abc}) and (\ref{3p_vertex}).

In the following, we neglect the subscript ``i" for simplicity.
Denote
\ea{{\label{alpha_beta_i}}
    \lrab{ \qsg(\eta_{\ast}, \vk_1) \qsg(\eta_{\ast}, \vk_2 ) }_{\alpha\beta 1} &\equiv - 2\, \textrm{Re} \lrsb{ \int^{\eta_{\ast} }_{{ -\infty^+} } d\eta_1\, \int^{\eta_1}_{{ -\infty^+}
        } d\eta_2\, \soev{0}{ \qsg(\eta_{\ast}, \vk_1) \qsg(\eta_{\ast}, \vk_2 ) \,  H_{\alpha}(\eta_1)\, H_{\beta}(\eta_2) }{0} } \,,\\
    \lrab{ \qsg(\eta_{\ast}, \vk_1) \qsg(\eta_{\ast}, \vk_2 ) }_{\alpha\beta 2} &\equiv  \int^{\eta_{\ast} }_{{ -\infty^-} } d\eta_1 \int^{\eta_{\ast}}_{{ -\infty^+} }
        d\eta_2\, \soev{0}{ H_{\alpha}(\eta_1) \, \qsg(\eta_{\ast}, \vk_1) \qsg(\eta_{\ast}, \vk_2 ) \,
         H_{\beta}(\eta_2) }{0} \,,
} where $H_\alpha$'s are given by (\ref{H_abc}) and
(\ref{3p_vertex}). Thus, the final loop-corrections read, \eq{
    \lrab{ \qsg(\eta_{\ast}, \vk_1) \qsg(\eta_{\ast}, \vk_2 ) }_{1-\textrm{loop}} =
    \sum_{\alpha,\beta =a,b,c} \sum_{i=1}^2 \lrab{ \qsg(\eta_{\ast}, \vk_1) \qsg(\eta_{\ast}, \vk_2 ) }_{\alpha\beta
    i} \,.
} It is useful to note that $\lrab{\cdots}_{\alpha\beta} =
\lrab{\cdots
 }_{\beta\alpha}^{\ast}$.

\subsection{Calculating the Loop Corrections}

Now we investigate the loop corrections. The calculation is
straightforward but rather lengthy. In this section we show the
explicit steps. Readers who are interested only in the final loop
contributions can go to the next subsection directly.

\subsubsection{Diagonal contributions}

\begin{figure}[h]
\centering
\begin{minipage}{0.7\textwidth}
  \centering
    \includegraphics[width=4.5cm]{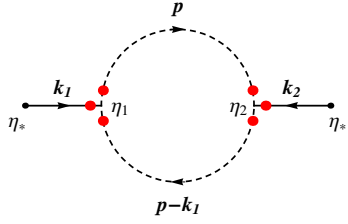}
    \caption{Diagrammatic representation of the (aa) loop contributions. A ``red dot" denotes derivative with respect to comoving time. It is important to note that the left and right end-points are labeled with $\eta_{\ast}$, while the left interaction vertex is labeled with $\eta_1$ and the right vertex is labeled with $\eta_2$.}
    \label{fig_aa1}
    \end{minipage}
\end{figure}

From (\ref{master}), the first integral reads,
    \eq{{\label{aa1}}
        \lrab{ \qsg(\eta_{\ast}, \vk_1) \qsg(\eta_{\ast}, \vk_2 ) }_{aa
        1} \equiv - 2\, \textrm{Re} \lrsb{ \int^{\eta_{\ast} }_{{ -\infty} } d\eta_1\, \int^{\eta_1}_{{ -\infty}
        } d\eta_2\, \soev{0}{ \qsg(\eta_{\ast}, \vk_1) \qsg(\eta_{\ast}, \vk_2 ) \,  H_{a}(\eta_1)\, H_{a}(\eta_2) }{0}
        } \,.
    }
The integral in the squared-bracket in (\ref{aa1}) is
    \ea{{\label{aa1_integral}}
      & (2\pi)^3 \delta^3(\vk_1 + \vk_2)  \;4\, g^2 \int^{\eta_{\ast} }_{{
-\infty} } d\eta_1\, \int^{\eta_1}_{{ -\infty}
        } d\eta_2\, \frac{1}{\eta_1 \eta_2} \, \od{}{\eta_1}G_{k_1}(\eta_{\ast}, \eta_1) \,
    \od{}{\eta_2} G_{k_1}(\eta_{\ast}, \eta_2) \\
    &\qquad\qquad \times \int \frac{d^3p}{(2\pi)^3}\, \frac{\textrm{d}^2}{\textrm{d}\eta_1 \textrm{d}\eta_2
    }F_{p}(\eta_1,\eta_2)\, \frac{\textrm{d}^2}{\textrm{d}\eta_1 \textrm{d}\eta_2
    }F_{|\vp -\vk_1|}(\eta_1,\eta_2) \,.
    }
Actually, this expression can be read from the ``Feynman-type"
diagramatic representation of the contribution, as shown in
Fig.\ref{fig_aa1}. The main differences from usual field theory is
that, here the perturbations are mostly ``derivatively-interacted",
rather than in usual field theory the interactions are described by
local products of fields. Moreover, in cosmological context, we are
interested in the expectation values of some arbitrary time rather
than scattering amplitudes. This is why the so-called ``in-in"
formalism has to be taken. The only thing we should note is the
\emph{order} among three ``time"
--- $\eta_\ast$, $\eta_1$ and $\eta_2$ between $\la 0|$ and $|0\ra$.
More precisely, in calculating the first integral of
(\ref{alpha_beta_i}), the order is (from left to right)
$\eta_{\ast}$-$\eta_1$-$\eta_2$, while in the second integral in
(\ref{alpha_beta_i}), the order is
$\eta_{1}$-$\eta_{\ast}$-$\eta_2$. Keeping these things in mind, we
are able to write the above explicit expressions from Feynman-type
diagrams directly.

Now, from (\ref{mode_function}) and (\ref{greenfunction}), we have
    \ea{
        \od{}{\eta_1}G_{k_1}(\eta_{\ast}, \eta_1) \,
    \od{}{\eta_2} G_{k_1}(\eta_{\ast}, \eta_2) &= \frac{c_s^4 H^4}{4k_1^2}\, \eta_1 \eta_2 \lrp{1 + ic_s k_1\eta_{\ast} }^2 e^{i c_s k_1 \lrp{ \eta_1 + \eta_2 - 2\eta_{\ast} }}  \,,\\
        \frac{\textrm{d}^2}{\textrm{d}\eta_1 \textrm{d}\eta_2
    } F_{p}(\eta_1,\eta_2)\, \frac{\textrm{d}^2}{\textrm{d}\eta_1 \textrm{d}\eta_2
    } F_{|\vk_2+\vp|}(\eta_1,\eta_2) &= \frac{c_s^4 H^4}{4} p\,
    \eta_1^2 \eta_2^2 \, \lrmb{ \vp + \vk_2 } e^{-i c_s \lrp{ p + \lrmb{\vp + \vk_2} }(\eta_1 - \eta_2)
    } \,.
    }
Thus, (\ref{aa1_integral}) can be written as
    \ea{
        & (2\pi)^3 \delta^3(\vk_1 + \vk_2)  \, g_a^2\, \frac{c_s^8
        H^8}{4k_1^2} \lrp{1 + ic_s k_1\eta_{\ast} }^2\, e^{- 2 i c_s k_1 \eta_{ \ast} }\; \int
        \frac{d^3p}{(2\pi)^3} \; p\; \lrmb{\vp - \vk_1} \\
        &\qquad \times \int_{-\infty}^{\eta_{\ast}} d\eta_1\;
        \eta_1^2\; e^{i c_s \lrp{k_1 - p -\lrmb{\vp - \vk_1} }\eta_1
        } \int_{-\infty}^{\eta_1} d\eta_2\; \eta_2^2\; e^{i c_s \lrp{ k_1 + p + \lrmb{ \vp - \vk_1 } }
        \eta_2} \,.
    }
The lower bound of both time integrals are $-\infty$, where the
integrants highly oscillate. However, we have to remind ourselves
that the lower bound should be chosen carefully in order to choose
the true vacuum (see Appendix A for a detailed discussion on this
problem and the issue of ``choosing vacuum"). Actually, here the
lower bound of $\eta_1$ integral and $\eta_2$ integral is the same,
$-\infty^+ \equiv -\infty(1-i\epsilon)$. Thus, the time integrals
can be evaluated without any ambiguity. For example, the integral
w.r.t. $\eta_2$ is evaluated as
    \eq{
        \int_{-\infty(1-i\epsilon)}^{\eta_1} d\eta_2\; \eta_2^2\; e^{i c_s \lrp{ k_1 + p + \lrmb{ \vp - \vk_1 } }
        \eta_2} = \frac{ \lrp{2i + 2c_s K \eta_1 - i c_s^2 K^2 \eta_1^2} e^{i c_s K \eta_1} }{c_s^3\, K^3} \,,
    }
where we write $K \equiv k_1 + p +\lrmb{ \vp - \vk_1 } $ for
simplicity. Thus, after performing the time integrals and taking the
real part, we finally get
    \ea{
        \lrab{ \qsg(\eta_{\ast}, \vk_1) \qsg(\eta_{\ast}, \vk_2 ) }_{aa
        1} =  -(2\pi)^3\, \delta^3(\vk_1 +\vk_2)\, g_a^2\, \frac{c_s^2
        H^8}{16 k_1^7} \int \frac{d^3p}{(2\pi)^3}\, \frac{p \lrmb{ \vp- \vk_1}}{K^3}\, \lrp{ a_1 K^2  + a_2 K \tilde{K}  + a_3 \tilde{K}^2 }
\,,
    }
where we have defined $\tilde{K} \equiv k_1 - p -\lrmb{ \vp - \vk_1
}$, and
    \ea{
        a_1 &\equiv 2\, \lrp{ 2 x_{\ast}^6 -5 x_{\ast}^2 -5 } \,,\\
        a_2 &\equiv  - \lrp{ 2 x_{\ast}^4 +5 x_{\ast}^2 +5 }\,,\\
        a_3 &\equiv  - \lrp{ 2 x_{\ast}^4 + x_{\ast}^2 +1 }\,,\\
    }
with $x_{\ast} \equiv -c_s k_1 \eta_{\ast}$ for short, note that $K
+\tilde{K} = 2k_1$.

The next step is to evaluate the momentum loop-integral:
    \eq{{\label{momentum_int_aa1}}
        I(k) \equiv \int \frac{d^3p}{(2\pi)^3}\,  \frac{p \lrmb{ \vp- \vk_1}}{K^3}\, \lrp{ a_1 K^2  + a_2 K \tilde{K}  + a_3 \tilde{K}^2 }  \,,
    }
remember that $ K \equiv k_1 + p +\lrmb{ \vp - \vk_1 }$ and
$\tilde{K} \equiv k_1 - p -\lrmb{ \vp - \vk_1 }$. Note that the
integral (\ref{momentum_int_aa1}) is IR finite, thus we do not need
to introduce infrared cutoff $\ell^{-1}$ here. It will be shown in
this work that loop corrections from the three-point vertices from
(\ref{3p_vertex}) are all IR finite, due to the
``derivatively-coupling" of the perturbations. The integrant in
(\ref{momentum_int_aa1}) has the structure $f(k_1,p,\lrmb{\vp
-\vk_1})$, in Appendix B, we describe the method to evaluate such
type of integrals. In particular, here the function
$f(k_1,p,\lrmb{\vp -\vk_1})$ is
    \[
        f(k_1,p,\lrmb{\vp -\vk_1}) =  \frac{p \lrmb{ \vp- \vk_1}}{K^3}\, \lrp{ a_1 K^2  + a_2 K \tilde{K}  + a_3 \tilde{K}^2
        } \,.
    \]
After a straightforward but tedious calculation, we get
    \eq{{\label{aa1_result}}
        \lrab{ \qsg(\eta_{\ast}, \vk_1) \qsg(\eta_{\ast}, \vk_2 ) }_{aa
        1} = (2\pi)^3 \delta^3(\vk_1 + \vk_2) P_{\sigma}(k_1)
        \frac{H_{\ast}^2}{\pi^2 \epsilon c_s^5} \lrp{ c_1 \ln k +
        \alpha_1
        } \,,
    }
where
    \ea{
        c_1 &= \frac{-4-4 x_*^2-25 x_*^4+x_*^6}{1920} \,,
    }
and $\alpha_1$ denotes the finite part of the momentum integral
together with the scheme-dependent UV renormalization constant,
which can be found in Appendix C .

Now let us move to the calculation of $(aa2)$ contribution. The
expression can also be read from fig.\ref{fig_aa1}:
    \ea{{\label{aa2}}
        &\quad\; \lrab{ \qsg(\vk_1,\eta_{\ast})\qsg(\vk_2,\eta_{\ast}) }_{aa
        2}\\
         &= (2\pi)^3 \delta^3(\vk_1 + \vk_2)  \;4\, g^2 \int^{\eta_{\ast} }_{{ -\infty} } d\eta_1\,
\int^{\eta_{\ast}}_{{ -\infty}
        } d\eta_2\, \frac{1}{\eta_1 \eta_2} \, \int \frac{d^3p}{(2\pi)^3}\, \od{}{\eta_1}G_{k_1}(\eta_1, \eta_{\ast}) \,
    \od{}{\eta_2} G_{k_1}(\eta_{\ast}, \eta_2) \\
    &\qquad\qquad \times \frac{\textrm{d}^2}{\textrm{d}\eta_1 \textrm{d}\eta_2
    }F_{p}(\eta_1,\eta_2)\, \frac{\textrm{d}^2}{\textrm{d}\eta_1 \textrm{d}\eta_2
    }F_{|\vp -\vk_1|}(\eta_1,\eta_2) \,.
    }
It is useful to note that, the only difference of integrant from
(aa1)-contribution is that, here in fig.\ref{fig_aa1}, the order of
time is $\eta_1 - \eta_\ast -\eta_2$. That is, what we need to do is
to simply change the first Green's function in (aa1)-case as
following
    \[
        \od{}{\eta_1}G_{k_1}( \eta_{\ast}, \eta_1) \qquad \Rightarrow \qquad \od{}{\eta_1}G_{k_1}(\eta_1,
        \eta_{\ast})\,,
    \]
to get the above result. Following the same strategy as above, we
finally get
    \eq{{\label{aa2_result}}
       \lrab{ \qsg(\vk_1,\eta_{\ast})\qsg(\vk_2,\eta_{\ast}) }_{aa2}
       = (2\pi)^3 \delta^3(\vk_1 + \vk_2)  \, P_{\sigma}(k_1)
        \frac{H^2}{\pi^2 \epsilon c_s^5} \lrp{ c_2 \ln k_1 +
        \alpha_{2}
        }\,,
    }
with
    \ea{
        c_{2} &= \frac{1}{384} x_*^4 \left(1+x_*^2\right) \,,
    }
and $\alpha_2$ can be found in Appendix C.

Collecting (\ref{aa1_result}) and (\ref{aa2_result}) together, we
get the contribution from (aa)-term
    \eq{{\label{aa_result}}
       \lrab{ \qsg(\vk_1,\eta_{\ast})\qsg(\vk_2,\eta_{\ast}) }_{aa}
       = (2\pi)^3 \delta^3(\vk_1 + \vk_2)  \, P_{\sigma}(k_1)
        \frac{H^2}{\pi^2 \epsilon c_s^5} \lrp{ c_{aa} \ln k_1 +
        \alpha_{aa}
        }\,,
    }
with $P_{\sigma}(k_1)$ is the (tree-level) power spectrum of
adiabatic mode $\qsg$, and
    \ea{
        c_{aa} & =c_1 +c_2 = \frac{1}{960}  \left(-2-2 x_*^2-10 x_*^4+3 x_*^6\right)
        \,.
    }

\begin{figure}[h]
\centering
\begin{minipage}{0.7\textwidth}
  \centering
    \includegraphics[width=4.5cm]{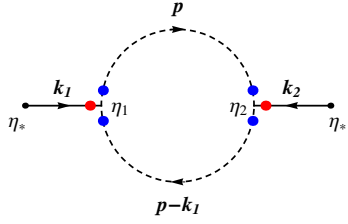}
    \caption{Diagrammatic representation of the loop corrections from (bb)-term. As before, a ``red dot" denotes derivative with respect to comoving time associated with the vertex, a ``blue dot" denotes a momentum factor associated with the line momentum. We may freely label the momentum flows into a vertex a ``$+$" sign and momentum flows out of a vertex a ``$-$" sign. The momentum factors of the same vertex are ``scalar-producted", e.g., the left vertex gives a factor $-\vp \cdot (\vp- \vk_1)$.}
    \label{fig_bb}
    \end{minipage}
\end{figure}

The contributions from (bb)-term can be read from Fig.\ref{fig_bb}.

    \ea{
        &\quad\; \lrab{ \qsg(\vk_1,\eta_{\ast})\qsg(\vk_2,\eta_{\ast}) }_{bb
        1} \\
         &= -(2\pi)^3 \delta^{3}(\vk_1 + \vk_2)\, 8\,
        g^2c_s^4 \times \textrm{Re} \int_{-\infty}^{\eta_{\ast}} d\eta_1
        \int_{-\infty}^{\eta_1} d\eta_2 \, \frac{1}{\eta_1 \eta_2}
        \int \frac{d^3p}{(2\pi)^3} \lrsb{\vp \cdot (\vp -\vk_1)}^2\\
        &\qquad\qquad \times
        \od{}{\eta_1}G_{k_1}(\eta_{\ast},\eta_1) \,
        \od{}{\eta_2}G_{k_1}(\eta_{\ast},\eta_2) \,
        F_p(\eta_1,\eta_2)\, F_{|\vp - \vk_1|}(\eta_1,\eta_2) \,.
    }
And
    \ea{
        \lrab{ \qsg(\vk_1,\eta_{\ast})\qsg(\vk_2,\eta_{\ast}) }_{bb
        2}  &= (2\pi)^3 \delta^{3}(\vk_1 + \vk_2)\, 4\,
        g^2c_s^4 \int_{-\infty}^{\eta_{\ast}} d\eta_1
        \int_{-\infty}^{\eta_{\ast}} d\eta_2 \, \frac{1}{\eta_1 \eta_2}
        \int \frac{d^3p}{(2\pi)^3} \lrsb{\vp \cdot (\vp -\vk_1)}^2\\
        &\qquad\qquad \times \od{}{\eta_1}G_{k_1}(\eta_1,
        \eta_{\ast}) \,
        \od{}{\eta_2}G_{k_1}(\eta_{\ast},\eta_2) \,
        F_p(\eta_1,\eta_2) \, F_{|\vp - \vk_1|}(\eta_1,\eta_2)
    }
After lengthy but straightforward calculation, we get
\eq{
    \lrab{ \qsg(\vk_1,\eta_{\ast})\qsg(\vk_2,\eta_{\ast}) }_{bb
        } = (2\pi)^3 \delta^3(\vk_1 + \vk_2)  \, P_{\sigma}(k_1)
        \frac{H^2}{\pi^2 \epsilon c_s^5} \lrp{ c_{bb} \ln k_1 +
        \alpha_{bb}}
} with
    \ea{
        c_{bb} &= \frac{ -273+27 x_*^2+90 x_*^4-8 x_*^6 }{3840} \,,
    }
agian $\alpha_{bb}$ is the finite part of the momentum integral and
can be found in Appendix C.

\begin{figure}[h]
\centering
\begin{minipage}{0.8\textwidth}
\centering
\begin{minipage}{0.4\textwidth}
  \centering
    \includegraphics[width=4.5cm]{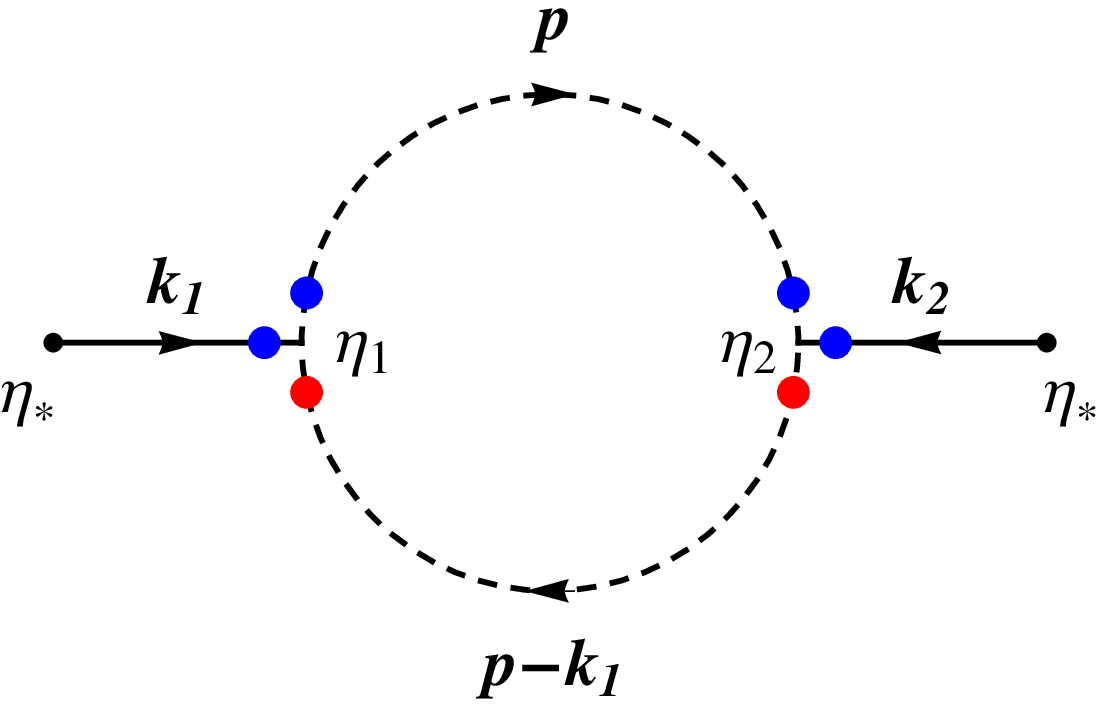}
    \end{minipage}
\begin{minipage}{0.4\textwidth}
  \centering
    \includegraphics[width=4.5cm]{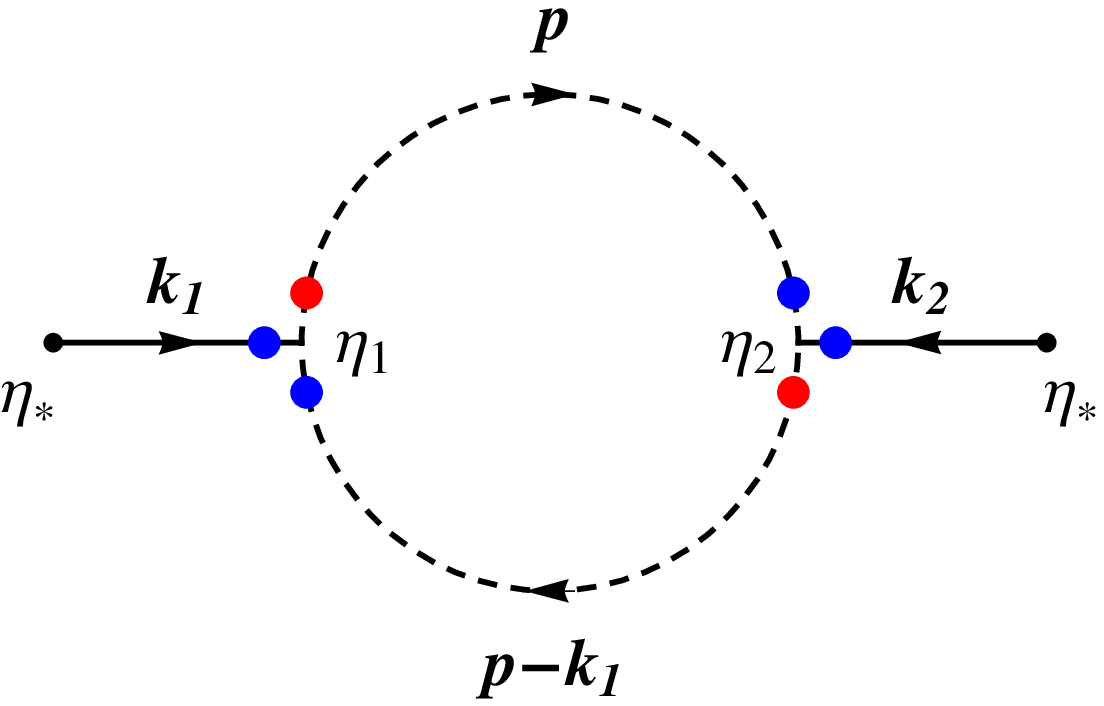}
    \end{minipage}
    \caption{Diagrammatic representation of the loop corrections from (cc)-contribution. As before, a red dot denotes derivative with respect to comoving time associated with the vertex, a blue dot denotes a momentum factor associated with the line momentum. Note that there are two types of diagrams from (cc)-contributions. }
    \label{fig_cc}
    \end{minipage}
\end{figure}

Contributions from (cc)-term can be read from Fig.\ref{fig_cc}.

    \ea{
        &\quad\; \lrab{ \qsg(\vk_1,\eta_{\ast})\qsg(\vk_2,\eta_{\ast}) }_{cc
        1}\\
         &= - (2\pi)^3 \delta^{3}(\vk_1 + \vk_2)\, 16\,
        g^2c_s^4 \times \textrm{Re} \int_{-\infty}^{\eta_{\ast}} d\eta_1
        \int_{-\infty}^{\eta_1} d\eta_2 \, \frac{1}{\eta_1 \eta_2}
        \int \frac{d^3p}{(2\pi)^3} \,  G_{k_1}(\eta_{\ast},\eta_1) \,
        G_{k_1}(\eta_{\ast},\eta_2) \\
        &\qquad\qquad \times \left\{ \lrp{\vp \cdot \vk_1}^2\,
        F_p(\eta_1,\eta_2)\, \frac{\textrm{d}^2}{\textrm{d}\eta_1 \textrm{d}\eta_2 } F_{|\vp -
        \vk_1|}(\eta_1,\eta_2) \right. \\
        &\qquad\qquad\qquad \left. - \lrp{\vp \cdot \vk_1}\, \lrsb{ \vk_1 \cdot(\vp -
        \vk_1) }\,
        \od{}{\eta_1} F_p(\eta_1,\eta_2)\, \od{}{\eta_2} F_{|\vp -
        \vk_1|}(\eta_1,\eta_2) \right\} \,.
    }
And
    \ea{
        &\quad\; \lrab{ \qsg(\vk_1,\eta_{\ast})\qsg(\vk_2,\eta_{\ast}) }_{cc
        2} \\
        &= (2\pi)^3 \delta^{3}(\vk_1 + \vk_2)\, 8\,
        g^2c_s^4 \int_{-\infty}^{\eta_{\ast}} d\eta_1
        \int_{-\infty}^{\eta_{\ast}} d\eta_2 \, \frac{1}{\eta_1 \eta_2}
        \int \frac{d^3p}{(2\pi)^3} \,  G_{k_1}(\eta_1, \eta_{\ast}) \,
        G_{k_1}(\eta_{\ast},\eta_2) \\
        &\qquad \times \left\{ \lrp{\vp \cdot \vk_1}^2\,
        F_p(\eta_1,\eta_2)\, \frac{\textrm{d}^2}{\textrm{d}\eta_1 \textrm{d}\eta_2 } F_{|\vp -
        \vk_1|}(\eta_1,\eta_2) \right.\\
        &\qquad\qquad \left. - \lrp{\vp \cdot \vk_1}\, \lrsb{ \vk_1 \cdot(\vp -
        \vk_1) }\,
        \od{}{\eta_1} F_p(\eta_1,\eta_2)\, \od{}{\eta_2} F_{|\vp -
        \vk_1|}(\eta_1,\eta_2) \right\} \,.
    }
Finally we get \eq{
    \lrab{ \qsg(\vk_1,\eta_{\ast})\qsg(\vk_2,\eta_{\ast}) }_{cc
        } = (2\pi)^3 \delta^3(\vk_1 + \vk_2)  \, P_{\sigma}(k_1)
        \frac{H^2}{\pi^2 \epsilon c_s^5} \lrp{ c_{cc} \ln k_1 +
        \alpha_{cc}}
} with
    \ea{
        c_{cc} &= \frac{1}{480}  \lrsb{-14+x_*^2 \left(-7+x_*^2\right) \left(2+x_*^2\right) } \,,\\
    }
and $\alpha_{cc}$ can be found in Appendix C.

\subsubsection{Off-diagonal contributions}

\begin{figure}[h]
\centering
\begin{minipage}{0.9\textwidth}
\centering
\begin{minipage}{0.3\textwidth}
  \centering
    \includegraphics[width=4.5cm]{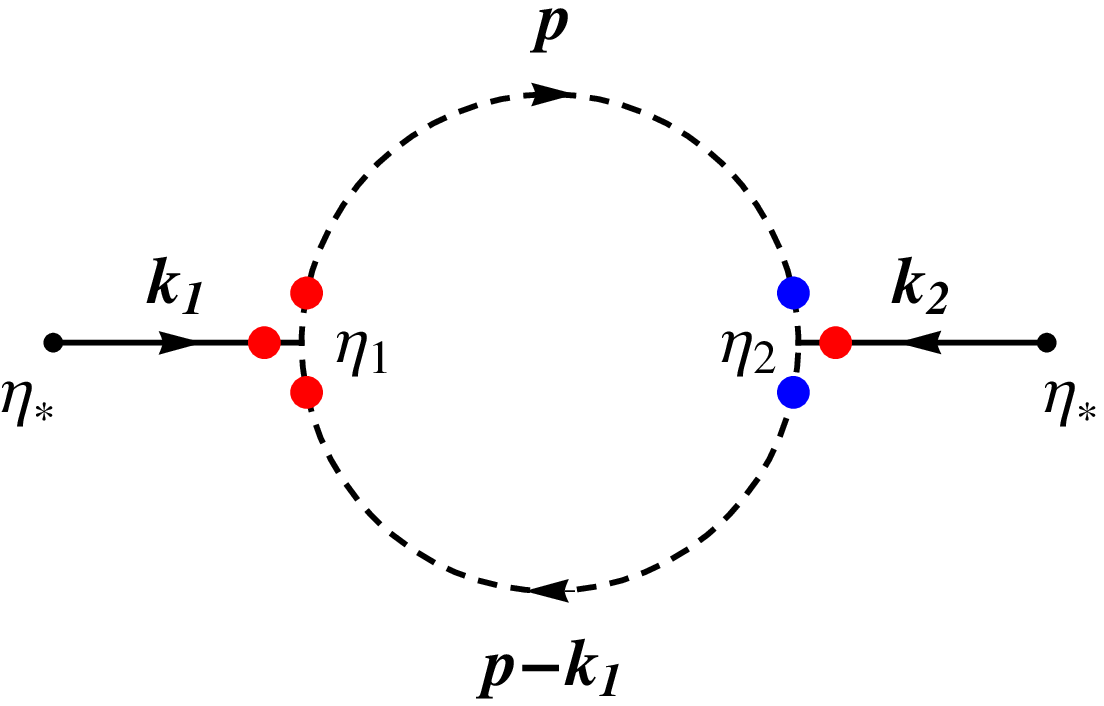}
    \end{minipage}
\begin{minipage}{0.3\textwidth}
  \centering
    \includegraphics[width=4.5cm]{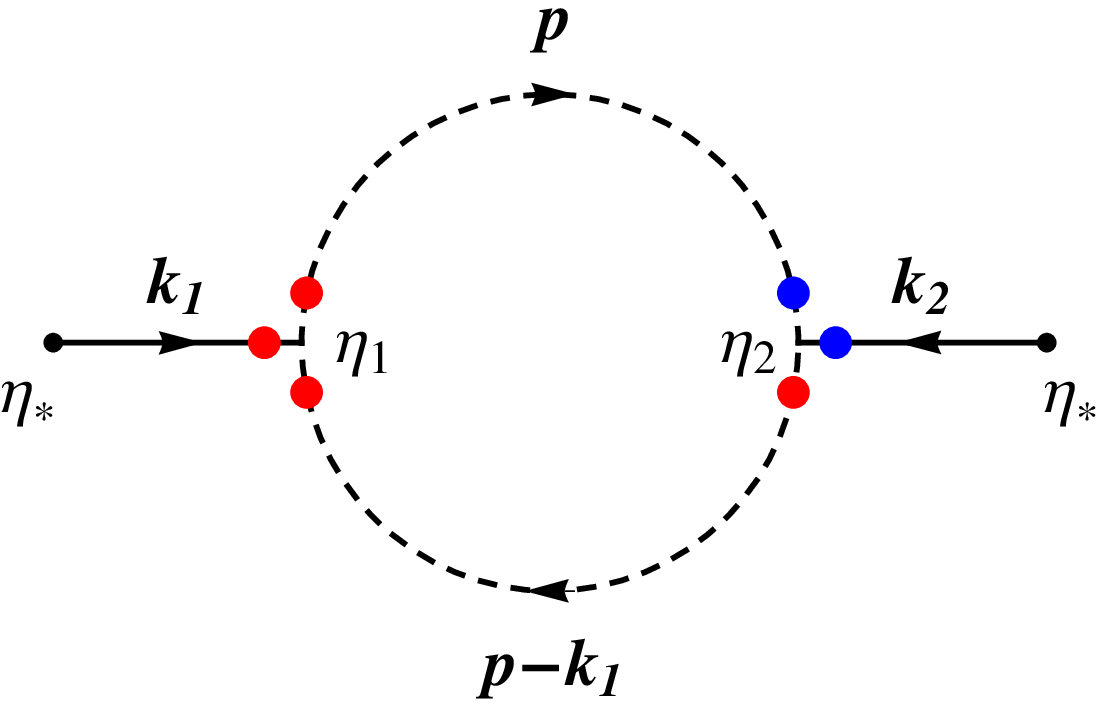}
    \end{minipage}
\begin{minipage}{0.3\textwidth}
  \centering
    \includegraphics[width=4.5cm]{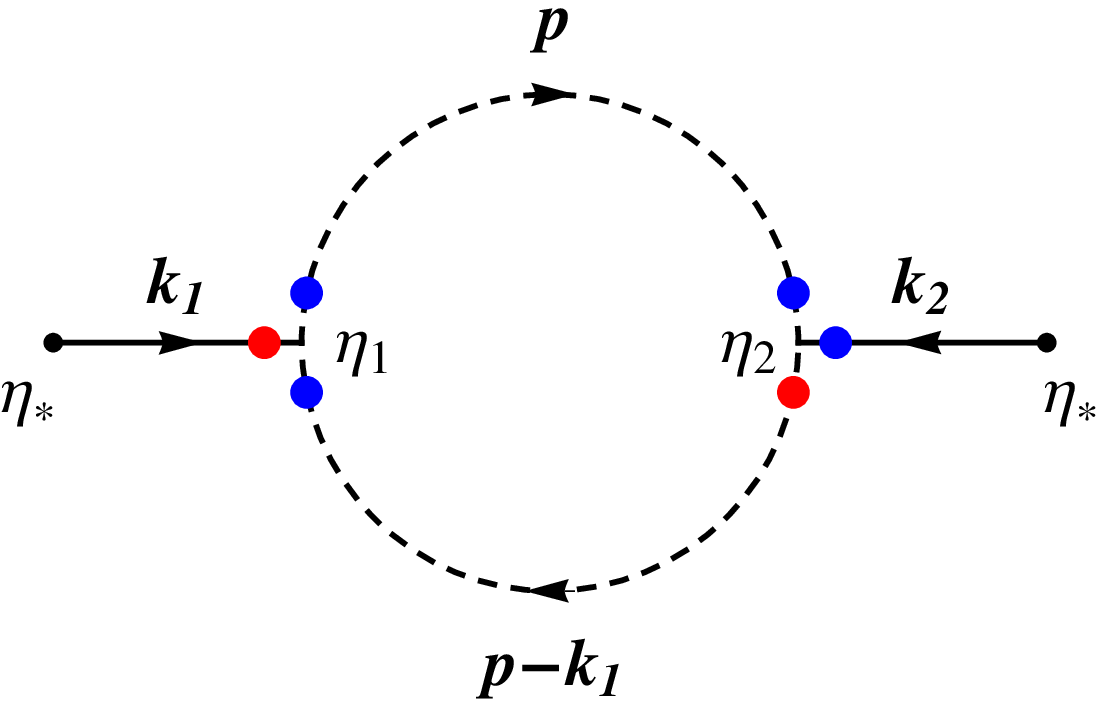}
    \end{minipage}
    \caption{Diagrammatic representations of off-diagonal contributions. From left to right are contributions from (ab), (ac) and (bc) combinations respectively. We do not show the (ba), (ca) and (cb) contributions explicitly.}
    \label{fig_off_diagonal}
    \end{minipage}
\end{figure}

In this work, we consider the one-loop two vertices contributions.
Besides the loop contributions from the same vertices, there are
contributions involving two different vertices, as shown in
Fig.\ref{fig_off_diagonal}. As described in the previous section,
their explicit expressions can be read from these Feynman-type
diagrammatic representations.

Contributions from (ab)-term are
    \ea{
        &\quad\; \lrab{ \qsg(\vk_1,\eta_{\ast})\qsg(\vk_2,\eta_{\ast}) }_{ab
        1}\\
         &= - (2\pi)^3 \delta^{3}(\vk_1 + \vk_2)\, 8
        g^2 c_s^2 \times  \textrm{Re} \int_{-\infty}^{\eta_{\ast}} d\eta_1
        \int_{-\infty}^{\eta_1} d\eta_2 \, \frac{1}{\eta_1 \eta_2}
        \int \frac{d^3p}{(2\pi)^3}\, \lrsb{ \vp \cdot (\vp - \vk_1) }  \\
        &\qquad\qquad \times \od{}{\eta_1} G_{k_1}(\eta_{\ast}, \eta_1)\,
        \od{}{\eta_2}G_{k_1}(\eta_{\ast},\eta_2) \,
        \od{}{\eta_1} F_{p}(\eta_1,\eta_2) \, \od{}{\eta_1} F_{|\vp
        -\vk_1|}(\eta_1,\eta_2) \,.
    }
And
    \ea{
        &\quad\; \lrab{ \qsg(\vk_1,\eta_{\ast})\qsg(\vk_2,\eta_{\ast}) }_{ab
        2}\\
         &= (2\pi)^3 \delta^{3}(\vk_1 + \vk_2)\, 4
        g^2 c_s^2 \int_{-\infty}^{\eta_{\ast}} d\eta_1
        \int_{-\infty}^{\eta_{\ast}} d\eta_2 \, \frac{1}{\eta_1 \eta_2}
        \int \frac{d^3p}{(2\pi)^3}\,\lrsb{ \vp \cdot (\vp - \vk_1) }   \\
        &\qquad\qquad \times \od{}{\eta_1} G_{k_1}(\eta_1, \eta_{\ast})\,
        \od{}{\eta_2}G_{k_1}(\eta_{\ast},\eta_2) \,
        \od{}{\eta_1} F_{p}(\eta_1,\eta_2) \, \od{}{\eta_1} F_{|\vp -\vk_1|}(\eta_1,\eta_2)
    }
The (ba) contributions can be also read from
Fig.\ref{fig_off_diagonal} which we do not write down explicitly.
Finally, we denote the contributions from (ab) and (ba) combinations
as: \eq{
    \lrab{ \qsg(\vk_1,\eta_{\ast})\qsg(\vk_2,\eta_{\ast}) }_{ab
        } = (2\pi)^3 \delta^3(\vk_1 + \vk_2)  \, P_{\sigma}(k_1)
        \frac{H^2}{\pi^2 \epsilon c_s^5} \lrp{ c_{ab} \ln k_1 +
        \alpha_{ab}}
} with
    \ea{
        c_{ab} &= \frac{ 4 x_*^6-10 x_*^4-11 x_*^2+49  }{3840} \,.
    }

Contributions from (ac)-term are:
    \ea{
        &\quad\; \lrab{ \qsg(\vk_1,\eta_{\ast})\qsg(\vk_2,\eta_{\ast}) }_{ac
        1}\\
         &= (2\pi)^3 \delta^{3}(\vk_1 + \vk_2)\, 16\,
        g^2 c_s^2 \times \textrm{Re} \int_{-\infty}^{\eta_{\ast}} d\eta_1
        \int_{-\infty}^{\eta_{1}} d\eta_2 \, \frac{1}{\eta_1 \eta_2}
        \int \frac{d^3p}{(2\pi)^3}\, \lrp{ \vp \cdot \vk_1 } \\
        &\qquad\qquad \times \od{}{\eta_1} G_{k_1}(\eta_{\ast}, \eta_1)\,
        G_{k_1}(\eta_{\ast},\eta_2) \,
        \od{}{\eta_1} F_{p}(\eta_1,\eta_2) \, \frac{\textrm{d}^2}{\textrm{d}\eta_1 \textrm{d}\eta_2 } F_{|\vp
        -\vk_1|}(\eta_1,\eta_2)  \,.
    }
And
    \ea{
        &\quad\; \lrab{ \qsg(\vk_1,\eta_{\ast})\qsg(\vk_2,\eta_{\ast}) }_{ac
        2} \\
        &= -(2\pi)^3 \delta^{3}(\vk_1 + \vk_2)\, 8\,
        g^2 c_s^2 \int_{-\infty}^{\eta_{\ast}} d\eta_1
        \int_{-\infty}^{\eta_{\ast}} d\eta_2 \, \frac{1}{\eta_1 \eta_2}
        \int \frac{d^3p}{(2\pi)^3}\, \lrp{ \vp \cdot \vk_1 } \\
        &\qquad\qquad \times \od{}{\eta_1} G_{k_1}(\eta_1, \eta_{\ast})\,
        G_{k_1}(\eta_{\ast},\eta_2) \,
        \od{}{\eta_1} F_{p}(\eta_1,\eta_2) \, \frac{\textrm{d}^2}{\textrm{d}\eta_1 \textrm{d}\eta_2 } F_{|\vp
        -\vk_1|}(\eta_1,\eta_2)  \,.
    }
Similarly, we denote the contributions from (ac) and (ca) as \eq{
    \lrab{ \qsg(\vk_1,\eta_{\ast})\qsg(\vk_2,\eta_{\ast}) }_{ac
        } = (2\pi)^3 \delta^3(\vk_1 + \vk_2)  \, P_{\sigma}(k_1)
        \frac{H^2}{\pi^2 \epsilon c_s^5} \lrp{ c_{ac} \ln k_1 +
        \alpha_{ac}}
} with
    \ea{
        c_{ac} &=  \frac{-28 x_*^6+6 x_*^4-23 x_*^2-23}{3840}
        \,.
    }

Contributions from (bc)-term are:
    \ea{
        &\quad\; \lrab{ \qsg(\vk_1,\eta_{\ast})\qsg(\vk_2,\eta_{\ast}) }_{bc
        1}\\
         &=  (2\pi)^3 \delta^{3}(\vk_1 + \vk_2)\, 16\,
        g^2 c_s^4 \times \textrm{Re} \int_{-\infty}^{\eta_{\ast}} d\eta_1
        \int_{-\infty}^{\eta_{1}} d\eta_2 \, \frac{1}{\eta_1 \eta_2}
        \int \frac{d^3p}{(2\pi)^3}\, \lrsb{ \vp \cdot (\vp -\vk_1) } \lrp{ \vp \cdot \vk_1
        }\,  \\
        &\qquad\qquad \times \od{}{\eta_1} G_{k_1}(\eta_{\ast}, \eta_1)\,
        G_{k_1}(\eta_{\ast},\eta_2) \,
        F_{p}(\eta_1,\eta_2) \, \od{}{\eta_2} F_{|\vp -\vk_1|}(\eta_1,\eta_2)
        \,.
    }
And
    \ea{{\label{bc2}}
        &\quad\; \lrab{ \qsg(\vk_1,\eta_{\ast})\qsg(\vk_2,\eta_{\ast}) }_{bc
        2} \\
        &= -(2\pi)^3 \delta^{3}(\vk_1 + \vk_2)\, 8\,
        g^2 c_s^4 \int_{-\infty}^{\eta_{\ast}} d\eta_1
        \int_{-\infty}^{\eta_{\ast}} d\eta_2 \, \frac{1}{\eta_1 \eta_2}
        \int \frac{d^3p}{(2\pi)^3}\, \lrsb{ \vp \cdot (\vp -\vk_1) } \lrp{ \vp \cdot \vk_1
        }\,  \\
        &\qquad\qquad \times \od{}{\eta_1} G_{k_1}( \eta_1, \eta_{\ast})\,
        G_{k_1}(\eta_{\ast},\eta_2) \,
        F_{p}(\eta_1,\eta_2) \, \od{}{\eta_2} F_{|\vp -\vk_1|}(\eta_1,\eta_2)
        \,.
    }
While the total contributions from (bc) and (cb) terms are,
\eq{{\label{bc}}
    \lrab{ \qsg(\vk_1,\eta_{\ast})\qsg(\vk_2,\eta_{\ast}) }_{bc
        } = (2\pi)^3 \delta^3(\vk_1 + \vk_2)  \, P_{\sigma}(k_1)
        \frac{H^2}{\pi^2 \epsilon c_s^5} \lrp{ c_{bc} \ln k_1 +
        \alpha_{bc}}
} with
    \ea{
        c_{bc} &=  \frac{ 12 x_*^6+66 x_*^4+147 x_*^2+227 }{3840}
        \,.
    }

\subsection{Final Results}

Collect (\ref{aa_result})-(\ref{bc}) together, the final one-loop
contributions from entropy mode to the adiabatic power spectrum is
\eq{
    \lrab{ \qsg(\eta_{\ast}, \vk_1) \qsg(\eta_{\ast}, \vk_2 )
    }_{1-\textrm{loop}} = (2\pi)^3 \delta^3(\vk_1 + \vk_2)
    P_{\sigma}(\vk_1) \frac{H_{\ast}^2}{\pi^2 \epsilon c_s^5} \lrp{ c \ln k_1 + \alpha
    } \,,
} with
    \ea{
        c &= \frac{1}{960} \left(18 x_*^4+5 x_*^2-35\right)  \,.
    }
Note that the above result should be supplemented with a
scheme-dependent UV renormalization constant. This result is
consistent with previous analysis
\cite{Weinberg:2005vy,Weinberg:2006ac,Sloth:2006az,Sloth:2006nu,Seery:2007we,Seery:2007wf,Dimastrogiovanni:2008af,Adshead:2008gk,Adshead:2009cb,Urakawa:2008rb}.
In this work, the loop corrections show no IR divergences. This is
mainly due to that in $c_s \ll 1$ limit, the leading-order
interaction vertices are dominated by ``derivatively-coupled" ones,
which enhance the IR safety. Moreover, our result agrees with the
bound obtained in
\cite{Shandera:2008ai,Leblond:2008gg,Cheung:2007st}, where
perturbation theory breaks down.

\section{Conclusion and Discussion}

In this work, we calculate the one-loop corrections to the power
spectrum of adiabatic mode from three-point adiabatic/entropy
cross-interactions, at the time of horizon exiting. As mentioned
before, in $c_s \ll 1$ models, there is enhancement of the effective
couplings for interaction vertices, thus contributions (higher-order
correlations functions and their loop-corrections) from
multi-vertices diagrams are in general cannot be neglected.
Precisely, one may expected the two-vertices one-loop contributions
and the one-vertex one-loop contributions are of the same order. We
find that, as the enhancement of non-gaussianities in models with
non-canonical kinetic terms due to the enhancement of small speed of
sound $c_s$, the loop corrections are also enhanced by both the
slow-roll parameter $\epsilon$ and $c_s$, more precisely, of order
$\sim \frac{H_{\ast}^2}{\epsilon c_s^5}$. Thus in general, the loop
corrections can be large. Moreover, we find that the loop
contributions are IR finite. Although in this work we shown these
``large" loop effect through explicit calculations, they can be
expected by a simple counting of the order of magnitude of the
interaction vertices. The point is that, in relatively large
parameter region we need to take into account the higher order loop
corrections when we discuss models with small $c_s$, which may be
important in the investigation of large non-Gaussianities.

There are several issues to be addressed. Firstly, the perturbations
of inflaton fields and thus the correlation functions themselves are
not observable directly. What is responsible for the CMB
anisotropies and LSS is the curvature perturbation $\zeta$. Thus we
should combine the perturbations of inflaton fields to yield the
perturbation of curvature perturbation long after horizon exiting,
e.g. by using $\delta N$ formalism
\cite{Starobinsky:1986fxa,Sasaki:1995aw,Lyth:2004gb}, as in
\cite{Seery:2007wf}. However, it is well-known that in multiple
field inflation models, the curvature perturbation is not conserved
on superhorizon scales, or in other words, there are also
cross-correlations among different modes after horizon exiting. Due
to these reasons, the complete analysis of loop corrections to the
curvature perturbation in multi-field models is a complicate task
which needs more subsequent work.

In this work, we focus on the three-point adiabatic/entroy
cross-interactions. While from (\ref{3rd_action}), in general, there
are adiabatic mode self-interactions. Moreover, in multiple-DBI
models, these various interaction-vertices give the same order of
contributions, both to non-gaussianties and loop corrections. A
complete analysis of loop corrections including contributions from
adiabatic loops and also the relations with observables (e.g. the
curvature perturbation and the CMB anisotropies) are also needed.

\bigskip

{\bf Acknowledgements}

We would like to thank Xiangdong Ji, Miao Li, Eugene Lim and Tao
Wang for useful discussions and comments. XG was supported by the
NSFC grant No.10535060/A050207, a NSFC group grant No.10821504 and
Ministry of Science and Technology 973 program under grant
No.2007CB815401. F. Xu acknowledges the hospitality from the TQHN
group at University of Maryland and the support from China's
Ministry of Education.


\appendix

\section{A Brief Review of ``In-in" Formalism}

\subsection{Preliminaries}

The ``in-in formalism" (also dubbed as ``Schwinger-Keldysh
formalism", or ``Closed-time path formalism")
\cite{Schwinger:1960qe,Calzetta:1986ey,Jordan:1986ug} is a
perturbative approach for solving the evolution of expectation
values over a finite time interval. It is therefore ideally suited
not only to backgrounds which do not admit an S-matrix description,
such as inflationary backgrounds.

In QFT applied in the calculation of S-matrix in particle physics,
the goal is to determine the amplitude for a state in the far past
$|\psi\ra$ to become some state $|\psi'\ra$ in the far future,
    \[
        \la \psi' | S |\psi \ra = \la \psi'(+\infty)
        |\psi(-\infty)\ra \,.
    \]
Here, conditions are imposed on the fields at both very early and
very late times. This can be done because that in Minkowski
spacetime, the states are assumed to be non-interacting at far past
and at far future, and thus are usually taken to be the free vacuum,
i.e., the vacuum of the free Hamiltonian $H_0$. The free vacuum are
assumed to be in ``one-to-one" correspondence with the true vacuum
of the whole interacting theory, as we adiabatically turn on and
turn off the interactions between $t=-\infty$ and $t=+\infty$.

While the physical situation we are considering here is quite
different. Instead of specifying the asymptotic conditions both in
the far past and far future, we develop a given state \emph{forward}
in time from a specified initial time, which can be chosen as the
beginning of inflation. In cosmological context, the initial state
are usually chosen as free vacuum, such as Bunch-Davis vacuum, since
at very early times when perturbation modes are deep inside the
Hubble horizon, according to the equivalence principle, the
interaction-picture fields should have the same form as in Minkowski
spacetime.

\subsection{``In" vacuum}

The Hamiltonian can be split into a free part and an interacting
part: $H=H_0+H_{\textrm{i}}$. The time-evolution operator in the
interacting picture is well-known
    \eq{{\label{U_op}}
        U(\eta_2,\eta_1) = \mathrm{T}\exp\left( -i\int_{\eta_1}^{\eta_2} dt' H_{\textrm{i}\I}(\eta')
        \right)\,,
    }
where subscript ``$\I$" denotes interaction-picture quantities,
$\textrm{T}$ is the time-ordering operator. Our present goal is to
relate the interacting vacuum at arbitrary time $|\Omega_\I(t)\ra$
with free vacuum $|0_\I\ra$ (e.g., Bunch-Davis vacuum). The trick is
standard. First we may expand $|\Omega_\I(\eta)\ra$ in terms of
eigenstates of free Hamiltonian $H_0$, $|\Omega_\I(\eta)\ra = \sum_n
|n_\I\ra\, \la n_\I| \Omega_\I(\eta)\ra$, then we evolve
$|\Omega_\I(\eta)\ra$ by using (\ref{U_op})
    \eq{{\label{vacuum_evo}}
        |\Omega_\I(\eta_2)\ra = U(\eta_2,\eta_1)
        |\Omega_\I(\eta_1)\ra = |0_\I\ra\,
        \la 0_\I| \Omega_\I\ra + \sum_{n\geq1} e^{+iE_n(\eta_2 - \eta_1)}\, |n_\I \ra\,
        \la n_\I| \Omega_\I(\eta_1)\ra \,.
    }
From (\ref{vacuum_evo}), we can immediately see that, if we choose
$\eta_2 = -\infty(1-i\epsilon)$, all excited states in
(\ref{vacuum_evo}) are suppressed. Thus we relate interacting vacuum
at $\eta= -\infty(1-i\epsilon)$ with free vacuum $|0\ra$ as
    \eq{
        |\Omega_\I(-\infty (1 - i\epsilon) )\ra = |0_\I \ra\,
        \la 0_\I| \Omega_\I \ra
    }
Thus, the interacting vacuum at arbitrary time $\eta$ is given by
    \ea{
        |\textrm{VAC},\textrm{in} \ra &\equiv |\Omega_\I(\eta)\ra =  U(\eta,-\infty(1-i\epsilon))  |\Omega_\I(-\infty(1-i\epsilon))\ra \\
        &= \mathrm{T}\exp\left( -i\int^{\eta}_{-\infty(1-i\epsilon)} d\eta'\, H_{\textrm{i}\I}(\eta')
        \right) |0_\I \ra\,
        \la 0_\I| \Omega_\I \ra \,.
    }

\subsection{Expectation values in ``in-in" formalism }

The expectation value of operator $\hat{\mathcal{O}}(\eta)$ at
arbitrary time $\eta$ is evaluated as
    \ea{{\label{in-in_ev}}
        \la \hat{\mathcal{O}}(\eta) \ra &\equiv \frac{ \la \textrm{VAC},\textrm{in} | \hat{\mathcal{O}}(\eta) |\textrm{VAC},\textrm{in}
        \ra }{ \la \textrm{VAC},\textrm{in} |\textrm{VAC},\textrm{in}
        \ra } \\
        &= \soev{ 0_\I }{
        \bar{ \mathrm{T}} \exp\left( i\int^{\eta}_{-\infty{ (1 + i\epsilon)} } d\eta' H_{1\I}(\eta')
        \right) \, \hat{\mathcal{O}}_\I(\eta) \, \mathrm{T}\exp\left( -i\int^{\eta}_{-\infty {( 1-i\epsilon)} } d\eta' H_{\textrm{i}\I}(\eta')
        \right) }{0_\I } \,,
    }
where $\bar{\textrm{T}}$ is the anti-time-ordering operator.

\begin{figure}[h]
    \centering
    \begin{minipage}{0.7\textwidth}
    \centering
        \includegraphics[width=8cm]{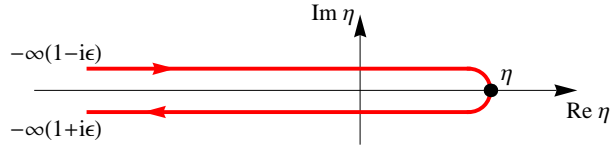}
    \end{minipage}
    \caption{Closed-time path in ``in-in" formalism.}
    \label{fig2_ctp}
    \end{figure}

For simplicity, we denote $-\infty(1-i\epsilon) \equiv -\infty^+ $
and $ -\infty(1+i\epsilon) \equiv -\infty^-$, since, e.g.,
$-\infty^+$ has a positive imaginary part. Now let us focus on the
time-order in (\ref{in-in_ev}). In standard S-matrix calculations,
operators between $\la 0|$ and $|0\ra$ are automatically
time-ordered. While in (\ref{in-in_ev}), from right to left, the
time starts from infinite past, or $-\infty^+$ precisely, to some
arbitrary time $\eta$ when the expectation value is evaluated, then
back to $-\infty^-$ again. This time-contour, which is shown in
Fig.\ref{fig2_ctp}, forms a closed-time path, so ``in-in" formalism
is sometimes called ``closed-time path" (CTP) formalism.

\subsection{Perturbation theory}

The starting point of perturbation theory is the free theory
two-point correlation functions. In canonical quantization
procedure, we write a scalar field as
    \eq{
        \phi_{\vk}(\eta) = u(k,\eta) a_{\vk} + u^{\ast}(k,\eta)
        a^{\dag}_{-\vk} \,,
    }
where $u(k,\eta)$ is the mode function for $\phi_{\vk}(\eta)$ (in
practice, $u_k(\eta)$ and $u^{\ast}_k(\eta)$ are two
linear-independent solutions of equation of motion for
$\phi_{\vk}(\eta)$, which are Wroskian normalized and satisfy some
initial or asymptotic conditions ).

The free two-point function takes the form
    \eq{{\label{2pf}}
        \soev{0}{ \phi_{\vk_1}(\eta_1) \phi_{\vk_2}(\eta_2) }{0}
        \equiv (2\pi)^3 \delta^3(\vk_1+\vk_2)  G_{k_1}(\eta_1,\eta_2)
        \,,
    }
with
    \eq{{\label{free_gf}}
        G_{k_1}(\eta_1,\eta_2) \equiv u_{k_1}(\eta_1)
        u^{\ast}_{k_1}(\eta_2) \,.
    }
In this work, we take (\ref{2pf}) and (\ref{free_gf}) as the
starting point.

Now Taylor expansion of (\ref{in-in_ev}) gives
    \itm{
        \item 0th-order
            \eq{{\label{ev_0th}}
                \lrab{ \hat{\mathcal{O}}(\eta) }^{(0)} = \la 0_\I | \hat{\mathcal{O}}_\I(\eta) |0_\I\ra \,.
                }
        \item 1st-order (one interaction vertex)
            \eq{
                \lrab{ \hat{\mathcal{O}}(\eta) }^{(1)} = 2\, \textrm{Re} \lrsb{ -i \, \int^{\eta}_{{ -\infty^+} } d\eta'\,
         \soev{0_\I}{ \hat{\mathcal{O}}_\I(t)\, H_{\textrm{i}\I}(t')  }{0_\I}
            }\,.
            }
         \item 2nd-order (two interaction vetices)
            \eq{{\label{ev_2nd}}
            \begin{aligned}
                \lrab{ \hat{\mathcal{O}}(\eta) }^{(2)} &= - 2\, \textrm{Re} \lrsb{ \int^{\eta}_{{ -\infty^+} } d\eta'\, \int^{\eta'}_{{ -\infty^+}
        } d\eta''\, \soev{0_\I}{ \hat{\mathcal{O}}_\I(\eta)\,  H_{\textrm{i}\I}(\eta')\, H_{\textrm{i}\I}(\eta'') }{0_\I} } \\
        &\qquad\qquad + \int^{\eta}_{{ -\infty^-} } d\eta' \int^{\eta}_{{ -\infty^+} }
        d\eta''\, \soev{0_\I}{ H_{\textrm{i}\I}(\eta') \, \hat{\mathcal{O}}_\I(\eta) \,
         H_{\textrm{i}\I}(\eta'') }{0_\I} \,.
         \end{aligned}
            }
    }
Here in this work, since we are considering one-loop contributions
from three-point interaction vertices, (\ref{ev_2nd}) is needed.

\subsection{Positive/negative-fields notation}

There is a formulation of ``in-in formalism" in terms of ``doubled
fields" in the literatures, mostly applied in the path-integral
formulation. The reason is that,
 the closed-time path (CTP) is imagined to circle the time axis in the complex $\eta$-plane (see
 fig.\ref{fig2_ctp}). Accordingly, we distinguish operator values from
 the upper and lower branches by labeling a ``$+$" on the upper
increasing-time contour and a ``$-$" on the lower decreasing-time
contour. Then it is convenient to introduce a new
``time-path-ordering operator" $\textrm{T}_{\textrm{p}}$, which
orders operator along the ``CTP-path", as shown in
Fig.\ref{fig2_ctp}. Obviously, $\textrm{T}_{\textrm{p}}$ acts on
$``+"$ fields as time-ordering operator $\textrm{T}$, and acts
``$-$" fields as anti-time-ordering operator $\bar{\textrm{T}}$.
Thus (\ref{in-in_ev}) can be recast into a more convenient and
familiar form
    \ea{{\label{doubled_form}}
        \la \hat{\mathcal{O}}(\eta) \ra
        &\equiv \soev{ 0 }{
        \mathrm{T}_{\textrm{p}}\, \exp\left( i\int^{\eta}_{-\infty } d\eta' H^-_{\textrm{i}}(\eta')
        \right) \, \hat{\mathcal{O}}_\I(\eta) \, \mathrm{T}_{\textrm{p}} \exp\left( -i\int^{\eta}_{-\infty  } d\eta' H^+_{\textrm{i}}(\eta')
        \right) }{0 } \\
        &= \soev{ 0 }{
        \mathrm{T}_{\textrm{p}}\,  \hat{\mathcal{O}}_\I(\eta) \, \exp\left( -i\int^{\eta}_{-\infty  } d\eta'
        \lrsb{H^+_{\textrm{i}}(\eta') -  H^-_{\textrm{i}}(\eta')}
        \right) }{0 } \,.
    }
Now, since we have two types of fields, the contractions between
different pairs yields four kinds of correlation functions,
    \eq{{\label{def_green_double}}
        \soev{0}{ \textrm{T}_{\textrm{p}}\, \phi^{\pm}_{\vk_1}(\eta_1) \phi^{\pm}_{\vk_2}(\eta_2)
        }{0} \equiv  (2\pi)^3
        \delta^3(\vk_1+\vk_2)G^{\pm\pm}_{k}(\eta_1,\eta_2) \,,
    }
where
    \ea{{\label{def_4_green}}
        G^{++}_{k}(\eta_1,\eta_2) &= G^>_k(\eta_1,\eta_2) \theta(\eta_1 - \eta_2) + G^<_k(\eta_1,\eta_2) \theta(\eta_2 - \eta_1)  \,,\\
        G^{--}_{k}(\eta_1,\eta_2) &= G^<_k(\eta_1,\eta_2) \theta(\eta_1 - \eta_2) + G^>_k(\eta_1,\eta_2) \theta(\eta_2 - \eta_1) \,,\\
        G^{+-}_{k}(\eta_1,\eta_2) &= G^<_k(\eta_1,\eta_2) \,,\\
        G^{-+}_{k}(\eta_1,\eta_2) &= G^>_k(\eta_1,\eta_2) \,,
    }
and
    \ea{ {\label{G^>G^<}}
        G^{>}_{k}(\eta_1,\eta_2) &\equiv u_k(\eta_1)
        u^{\ast}_k(\eta_2) \,,\\
        G^{<}_{k}(\eta_1,\eta_2) &\equiv u^{\ast}_k(\eta_1)
        u_k(\eta_2) \equiv \lrp{ G^{>}_{k}(\eta_1,\eta_2) }^{\ast}
        \equiv G^{>}_{k}(\eta_2,\eta_1) \,,
    }
are the vacuum Wightman functions. Now the perturbation calculations
are standard, after expanding the $\textrm{T}_{\textrm{p}}$-ordered
operator exponential in (\ref{doubled_form}).

The ``doubled-field" notation is a convenient ``book-marker" in
path-integral formulation of ``in-in formalism" in order to put
everything into ``one" exponential, while this notation itself is
not necessary in principle. Especially, in operator formalism, there
is no need to take this notation at all (see also
\cite{Adshead:2009cb} for a discussion on this point).

\section{Momentum Integral}

In this work, the momentum loop-integrals have the structure as
    \eq{{\label{p_integral}}
        I(k) \equiv \int \frac{d^3p}{(2\pi)^3}\, f(k,p,|\vp-\vk|) \,
    }
where $\vk$ is the momentum of external line, and $\vp$ is the loop
momentum. In this appendix, we briefly describe the methodology for
evaluating such type of integrals.

\begin{figure}[h]
  \centering
  \begin{minipage}{0.7\textwidth}
    \centering
    \includegraphics[width=5cm]{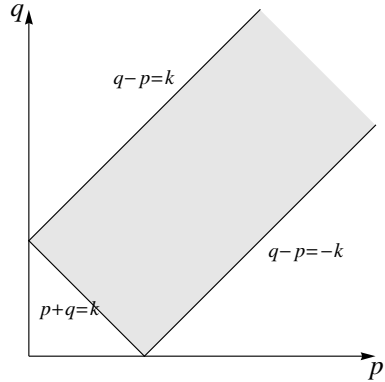}
    \caption{Integral region for variables $p$ and $q$.}
    \label{fig_region}
  \end{minipage}
\end{figure}

In general, one may use spherical coordinates $(p,\theta,\phi)$ to
calculate (\ref{p_integral}). However, it is more convenient to
introduce a new variable
    \eq{
        q \equiv  \lrmb{ \vp - \vk } = \sqrt{ k^2 + p^2 - 2kp\cos\theta
        } \,,
    }
and change coordinate $(p,\theta,\phi)$ into $(p,q,\phi)$. Note that
this can be done only when $k\neq 0$. The integral measure is simply
$ d^3p =  \frac{p q}{k} dp\, dq\, d\phi $. Thus, the original
integral (\ref{p_integral}) becomes
    \eq{{\label{pq_inte}}
        I(k) = \frac{1}{(2\pi)^2k} \int_0^{+\infty} dp \int_{\lrmb{p-k}}^{p+k} dq\; p\,q\, f(k,p,q) \,.
    }
The integral region for $p$ and $q$ is shown in
Fig.\ref{fig_region}.

One may use (\ref{pq_inte}) as the starting point. However, it can
be seen directly from Fig.\ref{fig_region} that sometimes it is more
convenient to use ``rotated" variables
    \ea{
        x &\equiv q+p \,,\qquad y\equiv q-p \,,\\
        p &= \frac{1}{2}(x-y)\,,\qquad q=\frac{1}{2}(x+y) \,,
    }
with integral measure $dp dq = \frac{1}{2}dx dy$. Finally, the
integral can be put into the form
    \eq{{\label{xy_form}}
        I(k) = \frac{1}{8(2\pi)^2k}\int_{k }^{\infty} dx\, \int_{-k}^k dy\;
        \lrp{x^2 - y^2}\, f\lrp{ k,\, \frac{1}{2}(x-y),\, \frac{1}{2}(x+y)
        } \,.
    }

\section{Finite parts}

For clarity, we collect the finite constant parts of the loop
integrals. \ea{ \alpha_1 &= \frac{-59-256 \ln 2-(59+256 \ln 2)
x_*^2+(464-1600 \ln
2) x_*^4+(-30+64 \ln 2) x_*^6}{122880} \,, \\
\alpha_{2} &=  \frac{\left(1+x_*^2\right) \left(331+30 (-21+40 \ln
2) x_*^4\right)}{460800}  \,. \\
\alpha_{aa} &=  \frac{439-3840 \ln2+(439-3840 \ln2) x_*^2-120
(-37+160 \ln2)
x_*^4+90 (-33+64 \ln2) x_*^6}{1843200}  \,,\\
\alpha_{bb} &= \frac{-2687-43680 \ln2+(34913+4320 \ln2) x_*^2+360
(-47+40 \ln2) x_*^4+(1410-1280 \ln2) x_*^6}{614400} \,,\\
\alpha_{cc} &=  \frac{5647-6720 \ln2+x_*^2 \left(5647-6720 \ln2+60
x_*^2 \left(23-40 \ln2+8 \ln2 x_*^2\right)\right)}{230400} \,, \\
\alpha_{ab} &= \frac{x_*^2 \left(30 x_*^2 \left(x_*^2 (11+32
\ln2)-452-80 \ln2\right)-11161-2640 \ln2\right)+4439+11760
\ln2}{921600}
\,, \\
\alpha_{ac} &=  \frac{-120 x_*^6 (28 \ln2-11)+20 x_*^4 (21+36
\ln2)+x_*^2 (601-2760 \ln2)+601-2760 \ln2}{460800}
        \,, \\
        \alpha_{bc} &= \frac{120 x_*^6 (12\ln2-19)+60 x_*^4 (77+132 \ln2)+x_*^2 (17640 \ln2-4849)-1249+27240 \ln2}{460800}
        \,.
 }
And \ea{
    \alpha &\equiv \frac{x_*^2 \left(-6 x_*^2 \left(x_*^2+22-18 \ln2\right)+347+30 \ln2\right)+137-210
    \ln2
    }{5760} \,.
}


\end{document}